# Experimental evaluation of a silicone oil as an oxidation inhibitor for Magnesium alloy under contact sliding at elevated temperatures

Yan Wang[a], Liangchi Zhang[b, c, d]*, Chuhan Wu[a]

[a] School of Mechanical and Manufacturing Engineering, The University of New South Wales, NSW 2052, Australia

[b] Shenzhen Key Laboratory of Cross-scale Manufacturing Mechanics, Southern University of Science and Technology, Shenzhen 518055, Guangdong, China

[c] SUSTech Institute for Manufacturing Innovation, Southern University of Science and Technology, Shenzhen 518055, Guangdong, China

[d] Department of Mechanics and Aerospace Engineering, Southern University of Science and Technology, Shenzhen 518055, Guangdong, China

*Corresponding author's E-mail: zhanglc@sustech.edu.cn

**Abstract**

This paper experimentally investigated the effects of silicone oil on the tribological behaviour of AZ31B/86CrMoV7 contact pair under a pin-on-disk configuration. A commercial silicone oil was used as an oxidation inhibiter for the AZ31B at elevated temperatures to 300 ºC. The wear mechanism of the contact pair was explored by SEM morphologies of worn surfaces. The analysis revealed that silicone oil can effectively minimize surface oxidation. Compared with dry sliding, silicone oil can considerably lower down the friction coefficient and specific wear rate. Under lubrication, the abrasive wear at a lower temperature can transit to the combined wear of abrasion, fatigue and adhesion when the temperature rises.

# 1 Introduction

Magnesium (Mg) and its alloys are promising structural materials for automotive, electronics and aerospace applications due to their lightweight and a good combination of engineering properties [1]. Unfortunately, at room temperature (RT), the poor material ductility of Mg alloys, which is the result of lacking slip systems inherent to the hexagonal close-packed (HCP) crystal structure [2], limits their manufacturability. Additionally, the nonlinear strain hardening behaviour at low ambient temperatures makes the fabrication of Mg alloys even harder [3]. As such, based on the structural metallurgy of Mg alloys, several attempts have been made to improve the formability to enhance their engineering applications. Investigations have demonstrated that an elevated ambient temperature not only improves the ductility by activating the slip systems of Mg alloys but also refines the grain size [4-6]. Therefore, commercial Mg alloy components are mainly produced in open atmospheres with elevated temperatures between 200 and 350 ºC [7]. This temperature window is conducive to the plasticity and ductility as well as to the minimization of the hardening effect of Mg alloys [8-10].

However, the elevated temperature inevitably results in the thermal oxidation of Mg alloys [11]. The oxidation in the open atmosphere presents different functions to the substrate under different temperatures, i.e., protective or unprotective, which makes the forming production of Mg alloy components difficult. The thermal oxides formed at a low temperature were protective and conducive to resisting the surface wear. For instance, a mechanical mixing layer with high hardness and stiffness can emerge in the regime around 200 ºC, which reduces successive wear [12, 13]. However, thick oxides formed at a high temperature are unfavourable because they are usually loose in structure [14, 15], such as the cauliflower-like morphologies of thick oxide nodules [16] which are prone to cracking. Moreover, unexpected oxide nodules were observed both inside and outside the wear tracks under dry contact sliding [17]. This became more significant when the contact sliding was at a moderate/high speed and load during which the high friction brought about high-temperature rises. These oxide nodules result in a significant wear rate and lead to the adhesion of Mg alloys to counterpart surfaces. Hence, in the forming production of Mg Alloy components at high temperatures, such surface oxidation must be eliminated.

To this end, oxidation inhibition has been tried out. Generally, the suppression mechanism can be isolating Mg alloys from the environment, e.g., applying protective coatings [18, 19] or adding reactive elements (RE) to the alloy to retard or inhibit possible oxidation [20, 21]. However, the coating requires complicated processing technologies with expensive chemical reagents, some of which, e.g., fluorine-containing chemicals, are not environmentally friendly [22]. Additionally, it is neither an effective approach to eliminate the manufacturing-induced oxidation nor suitable for large-scale industrial production [16]. While adding certain RE can reduce the extent of oxidation, it is hard to meet the industrial requirement for the fabrication of a specific class of products. A promising method is probably to apply lubricant to minimize the direct contact between the surfaces of an Mg alloy workpiece and the production atmosphere to suppress surface oxidation. An additional advantage of this approach is that applying lubricant is already essential to the control of many manufacturing processes for metal products [23, 24], such as the control of friction.

At elevated temperatures, silicone oil has been widely used in lubrication because of its favourable characteristics, e.g., eco-friendly trait, film strip formations and high lubricity [25-27]. Most investigations on silicone oil have been on its inherent thermal stability in production [28]. A systematic investigation on its role in suppressing surface oxidation and in determining interface friction and wear of Mg alloys, however, is particularly lacking.

This study aims to experimentally investigate the effect of silicone oil on the suppression of surface oxidation of a typical Mg alloy (AZ31B) subjected to contact sliding at various temperatures in a pin-on-disk testing chamber. The wear mechanisms were explored with the aid of microscopy analysis of the worn surfaces.

## 2 Materials and methods

### 2.1 Materials

A chemically inert silicone oil (Huber P20.275.50), consisting of polydimethylsiloxane (PDMS), was selected as the base lubricant for the tribo-tests to inhibit surface oxidation of Mg alloy AZ31B under contact sliding at various temperatures. The selected silicone oil is thermally stable due to the strength of the Si–O bond and can form a film over the AZ31B surface to minimize the surface exposure to the air [29]. The oil is safe to the Mg alloy, can be recycled, and is free from sulfur, chlorine, and other substances that may be harmful to the environment. Table 1 lists some primary properties of the oil.

**Table 1** Main properties of the base lubricant

| Properties | Value |
|---|---|
| Substance | PDMS |
| Flash point (°C) | >300 |
| Ignition temperature (°C) | 450 |
| Kinematic viscosity ($mm^2 s^{-1}$ at 25 °C) | 50 |
| Density (g/cm³ at 20 °C) | 0.96 |
| Heat conductivity (Wm⁻¹K⁻¹ at 50 °C) | 0.14 |
| Thermal expansion coefficient ($10^{-5}K^{-1}$) | 95 |

The contact sliding tests were conducted on a pin-on-disc tester (CETR UMT 100, Bruker, USA) with AZ31B as the disc samples (diameter 40 mm and thickness 4 mm). The chemical compositions of AZ31B are shown in Table 2. The pin for testing was made from a much harder metal, 86CrMoV7, which is a common roll material in rolling production. The pin was machined to be cylindrical with a flat tip and a diameter of 2 mm. In the preparation, the pin and disc samples were sequentially ground with coarse (320 grit) to fine (1200 grit) SiC papers and polished by 0.1 μm alumina power slurry before a 20 mins ultrasonic cleaning [30]. A smooth surface finish was obtained with an average surface roughness of Ra ≤ 0.5 μm. Before a microstructural investigation, a prepared sample was etched by an acetic picric solution (5 mL acetic acid + 6 g picric acid + 100 mL ethyl alcohol + 10 mL $H_2O$) to expose the microstructures [31]. Under an optical microscope, the microstructure of an untested AZ31B disc was uneven, consisting of small grains interspersed between large grains. The average grain size of the untested disc was around 24 μm, as shown in Fig.1.

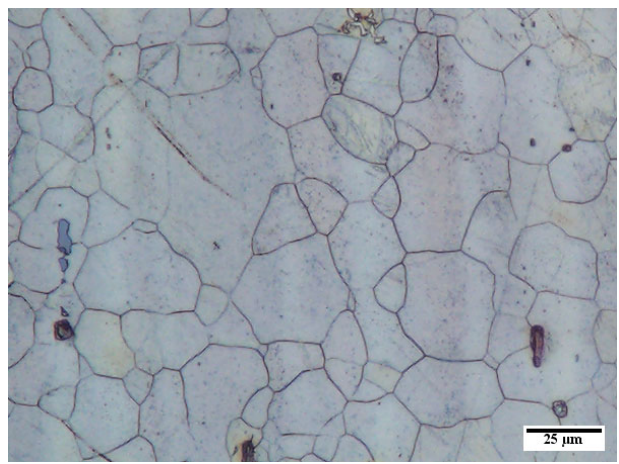

**Fig. 1** Optical microstructure of an AZ31B disc before sliding test

**Table 2** Chemical compositions of AZ31B (wt. %)

| Elements | Al | Zn | Ca | Mn | Fe | Cu | Mg |
|---|---|---|---|---|---|---|---|
| Composition | 2.5~3.5 | 0.6~1.4 | 0.002 | 0.2~0.4 | 0.005 | 0.03 | balance |

**2.2 Testing conditions**

As shown in Fig. 2, the anti-oxidation and anti-friction properties of the silicone oil lubricant were evaluated on the heating chamber of the CETR UMT 100 tribometer mentioned before. In a test, the disc sample was fixed in the lubricant container and the pin was fixed onto a load cell which provided a force vertical to the disc surface and that tangential to the surface in the sliding direction. To simulate the conditions in an industrial production rolling process of Mg alloy, the sliding speed and normal load were fixed at 5 mm/s and 5 N [32, 33], respectively. Such experimental conditions can also avoid possible excessive frictional heating [34]. Before a test, the ambient temperature was controlled by the heating chamber to have reached a desired temperature (one in the range of RT to 300 ºC) and maintained stable. To isolate surface changes prior to a contact sliding test, all sample discs were not embedded in the holder until the desired temperature stabilized. A thermal imaging camera and a thermocouple were used to monitor the actual temperature during the experiment. Each test was repeated at least 3 times under a nominally identical condition with a duration of 30 mins for a test. The wear debris was collected from the lubricant after a test was completed.

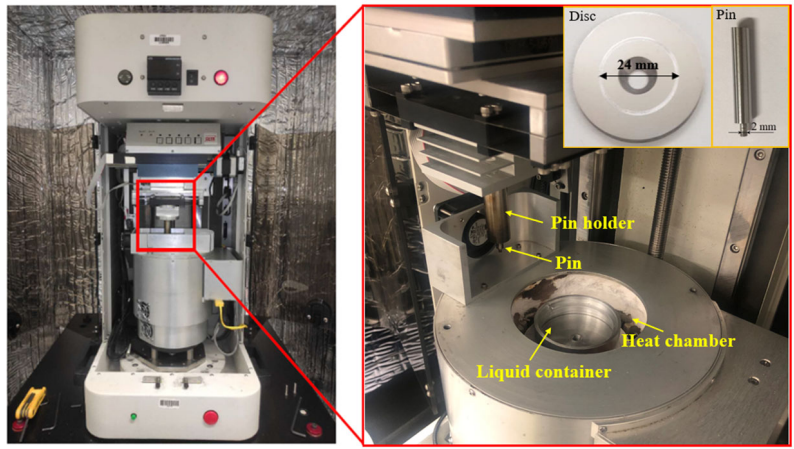

**Fig. 2** CETR UMT 100 tribometer with a heating chamber.

**2.3 Measurements**

The surface topography of the tested disc samples and the pin were examined and measured on an Olympus DSX 510 microscope [35]. The specific wear rate was calculated by Archard's law [36]. Before a measurement, the tested disc sample was ultrasonically cleaned to remove silicone oil residues. An acetone with a low concentration of sodium hydroxide additive was used for better removal results. The disc sample surface, both inside and outside the wear track, was examined by using a Keyence VXX200 laser microscope at 150x magnification. In addition, the disc and pin hardness was measured by a Struers Durascan-80 hardness tester. An FEI Nova Nano SEM 450 scanning electron microscope (SEM) at 5keV was used to visualize the wear track morphology. A Bruker SDD-EDS detector, configured to the SEM 450, was used for the element analysis.

**3 Results and discussion**

**3.1 Properties of disc samples**

The AZ31B disc surfaces were initially examined before the tribology tests. Typically, the freshly polished disc surfaces were smooth but with some polishing defects [17]. Two sets of discs were placed in the air and in the silicone oil, respectively, under RT and 300 ºC for 30 mins. As shown in Figs. 3a and c, both sets of the disc surfaces under RT present similar topography, e.g., smooth albeit defective. After the exposure in the air at 300 ºC for 30 mins, evenly distributed dark spots appeared on the disc surface. The laser intensity (Fig. 4a) and height map (Fig. 4b) showed that these spots show a cauliflower-like appearance. Previous investigations on such kind of spot appearance have demonstrated that they are oxide nodules [11, 16, 17, 37]. On the other hand, the samples covered with the silicone oil under the same temperature did not show noticeable surface changes, as shown in Fig. 3d. However, at a high temperature, the surfaces of both sets of samples presented similar crack patterns as shown in Figs. 4a and 3d.

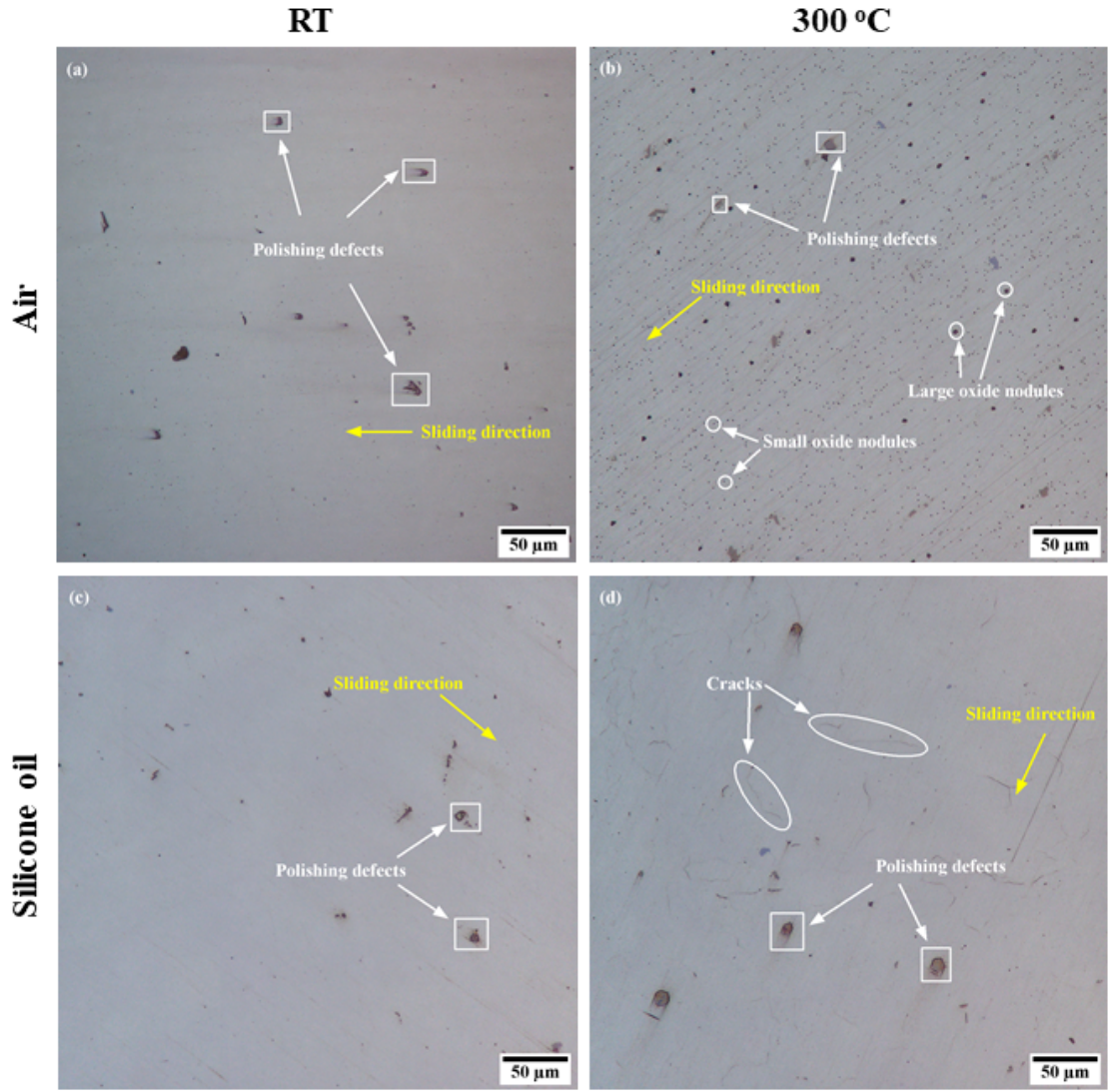

**Fig. 3** Disc surfaces placed in the air and silicone oil, respectively, under RT and 300 ºC for 30 mins.

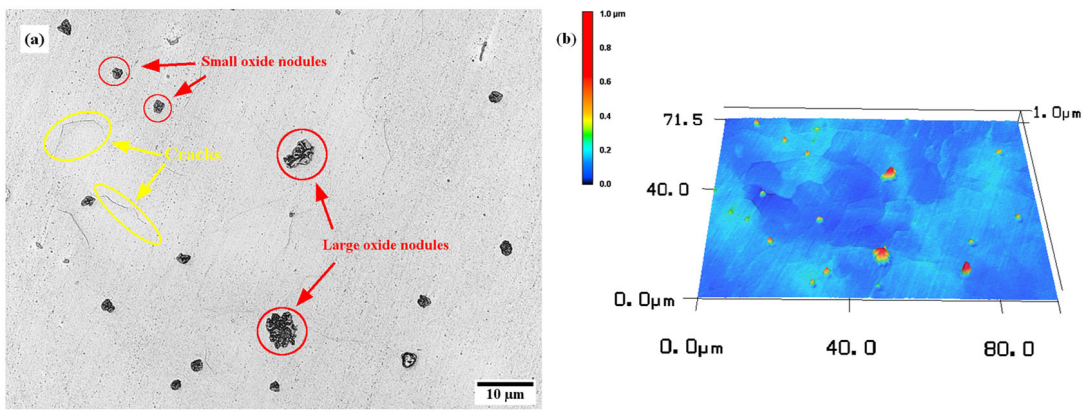

**Fig. 4** Oxide nodules on the AZ31B disc surface after the exposure in the air for 30 mins at 300ºC. (a) laser intensity, (b) 3D heightmap.

The discrepancy on the surface morphology in Fig. 3 may imply a difference in the surface chemical compositions. A comparative analysis on surface chemical compositions was carried out using the Bruker SDD-EDS detector for further investigation (Fig. 5 and Table 3). Figs. 5a and b are EDS spectra results of disc surfaces processed for 30 mins without and with silicone oil protection, respectively. The processing temperatures are denoted in different colours, with red for RT and blue for 300 ºC, respectively. Fig. 5a shows Mg, Al, Zn, and Mn peaks, which are expected chemical compositions of the AZ31B material in Table 2, whereas the C and O peaks confirm the oxidation of disc surface in the air. The intensity of the C and O peaks indicates oxidation degree [38]. Similar peaks were also detected on disc surfaces with silicone oil protection, except that the Si peak was originated from silicone oil, as shown in Fig. 5b.

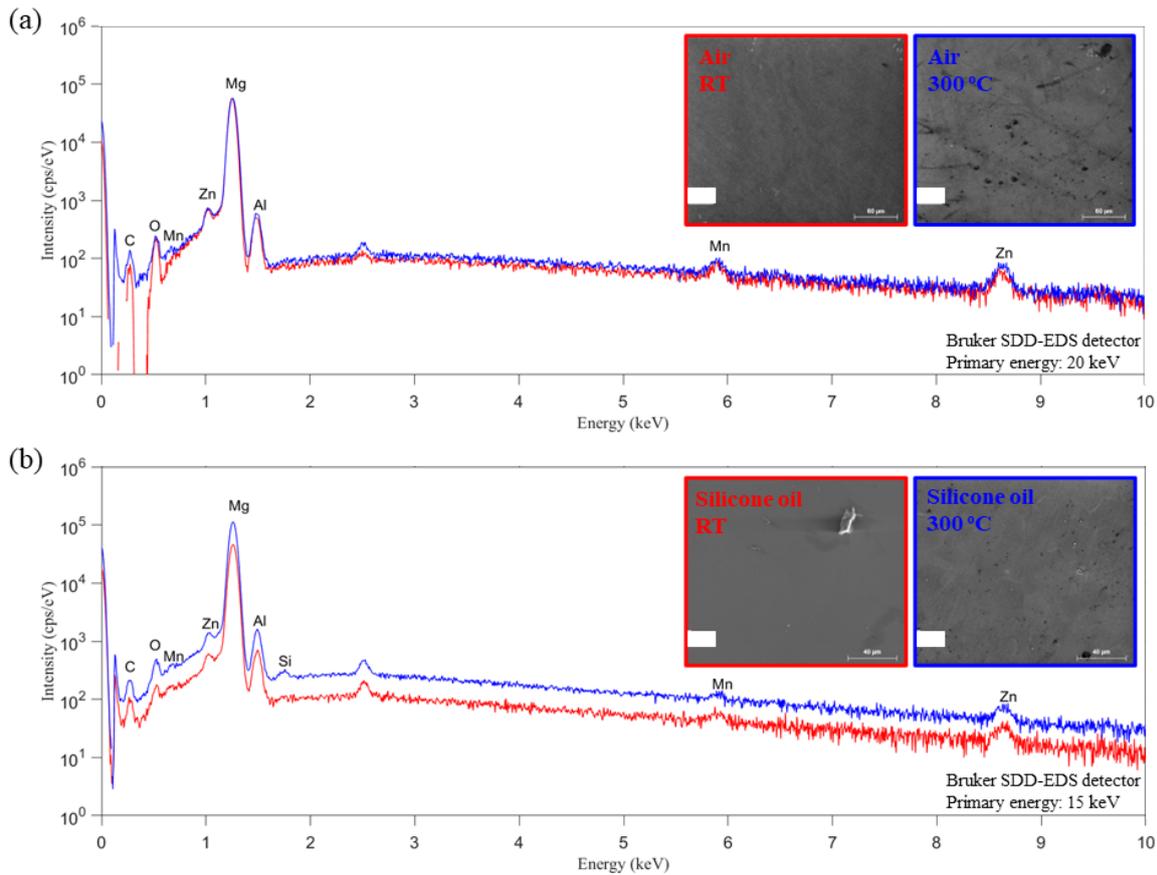

**Fig. 5** Energy dispersive spectrometry (EDS) spectra of disc surface after the exposure in (a) air, and (b) silicone oil for 30 mins at RT and 300ºC, respectively.

Table 3 lists quantitative results of 7 main elements corresponding to the EDS spectra. From RT to 300 °C in the air, the C and O percentage increased to 8.54 at. % and 1.63 at. %, respectively. This increase could be attribute to the emergence of oxide nodules. This result is consistent with the previous analysis for the oxide nodules formed by Mg and C elements on the non-protective oxide film [17]. Similar results can also be found in the study by Czerwinski et al. [11, 16], Arrabal et al. [15], and Tan et al. [37]. In contrast, with protection by silicone oil, there is only a small increase in the percentage of C from RT to 300 ºC. Moreover, the silicone oil has a chemical formula of $(C_2H_6OSi)n$ [29], and hence such a small increase in C and O may also be due to the silicone oil residues, as evidenced by the trace of Si [39]. With all these, one can conclude that silicone oil does protect the Mg alloy from the formation of oxide nodules at high temperatures.

Table 3 EDS analysis on disc surface (at. %)

| Condition | | Mg | Al | Zn | Mn | C | O | Si | Total |
|---|---|---|---|---|---|---|---|---|---|
| Air | RT | 91.94 | 2.04 | 0.35 | 0.14 | 4.08 | 1.45 | 0 | 100 |
| | 300 °C | 87.12 | 2.05 | 0.51 | 0.15 | 8.54 | 1.63 | 0 | 100 |
| Silicone oil | RT | 91.58 | 2.37 | 0.58 | 0.15 | 4.93 | 0.39 | 0 | 100 |
| | 300 °C | 88.53 | 2.44 | 0.51 | 0.11 | 6.2 | 2.1 | 0.11 | 100 |

An Mg alloy at high temperatures in the air often shows surface cracking and formation of oxide nodules due to oxidation and stress-concentration [40]. In this process, the material usually experiences an outward diffusion, leading to the formation of voids at the interface between the MgO layer and the adherent alloy grain boundary [41]. The voids can grow and react with the interstitials, causing stress-concentration and the generation of ridges and nodules in the oxide layer [11, 16]. Meanwhile, the $Mg^{2+}$ travels through voids and reacts with oxygen in the air, eventually resulting in cracked ridges and oxide nodules with a cauliflower morphology, as schematically demonstrated in Fig. 6. With the application of the silicone oil, however, the direct contact between $Mg^{2+}$ and oxygen in the air did not occur, and hence prevented the emergence of oxide nodules. Nevertheless, cracks can still be noted on the disc surfaces due to an accelerated ion diffusion and void formation at 300 °C, as shown in Fig.3d and 4a.

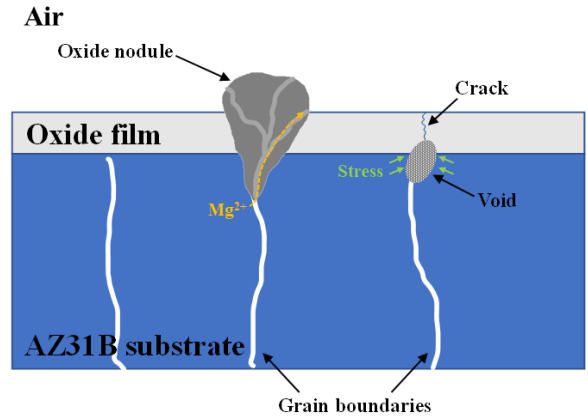

**Fig. 6** Schematic diagram of the oxide nodule and crack

The surface hardness of the AZ31B disc was measured to investigate its formability [42, 43]. At different temperatures, the disc samples were pre-heated for 30 mins in the open atmosphere and silicone oil, respectively, before being cleaned (see Section 2.1) to remove the residue lubricant. There were 21 mapping points for the indentations on each disc surface by a Vickers pyramid indenter with a load of 1 kgf and a duration of 15 s as per ASTM specifications. After the indentation test, the average diagonal indentation length was measured to calculate the surface hardness. Fig. 7 shows the obtained surface hardness under different pre-heating temperatures. Apparently, the thermal softening effects, which are dependent on the refined grain size at elevated temperatures [44], on the surface hardness can be noted for both the dry and lubricated cases. Moreover, it seems that the surface hardness for the lubricated case is slightly higher than that without lubrication because the surface oxidation in the latter can also alter the surface properties as the temperature increases.

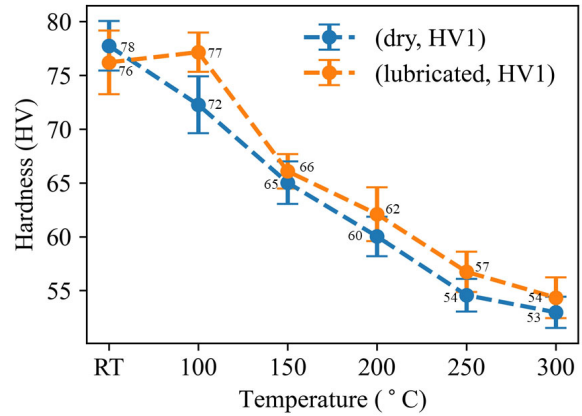

**Fig. 7** AZ31B disc surface hardness after the pre-heating for 30 mins in the air and silicone oil

### 3.2 Lubrication performance of the silicone oil

To understand the lubrication performance of silicone oil at different ambient temperatures, sliding tests with a contact pair of AZ31B/86CrMoV7 were conducted. The measured instant friction coefficients were plotted in Fig. 8a. The coefficient of friction (COF) is stable at a lower ambient temperature, i.e., 100 °C, but it experiences larger fluctuations at an elevated temperature. A possible reason for this may be the formation of a stable lubricant film at the contact interface under the lower temperature [17, 45]. However, the significant drop in lubricant viscosity, i.e., a reduction from 50 $mm^2$/s at RT to less than 10 $mm^2$/s at and beyond 140 °C, lead to an extremely thin oil film or even the breakdown of the lubricating film at higher temperatures [46-48]. As such, more direct contacts at the interface occurred, which resulted in larger COF values and fluctuations. Despite this, it still has a lubricating effect and can protect the contact surface to a certain extent in comparison to a dry sliding test. Additionally, the COF tends to decrease as the lubricated sliding test proceeds at an elevated temperature, e.g., the reduction from 0.35 to 0.15 at 300 °C. This may be due to the oxidation inhibition of the silicone oil. The oil inhibits the successive oxidation on the wear track after the initial material removal during sliding [49]. In contrast, the average COF for the dry sliding significantly increases from 0.32 to 0.70 when the ambient temperature increases from RT to 300 °C due to the emergence of oxide nodules on the wear track at a high temperature, e.g., 300 °C. In fact, these oxide nodules can accelerate the breakdown of the oxide layer and lead to a rough contact surface [50]. Compared with the dry sliding contact, a lower variation range, i.e., from 0.15 to 0.32, can be noted for the average COF with the silicone oil, as shown in Fig. 8b.

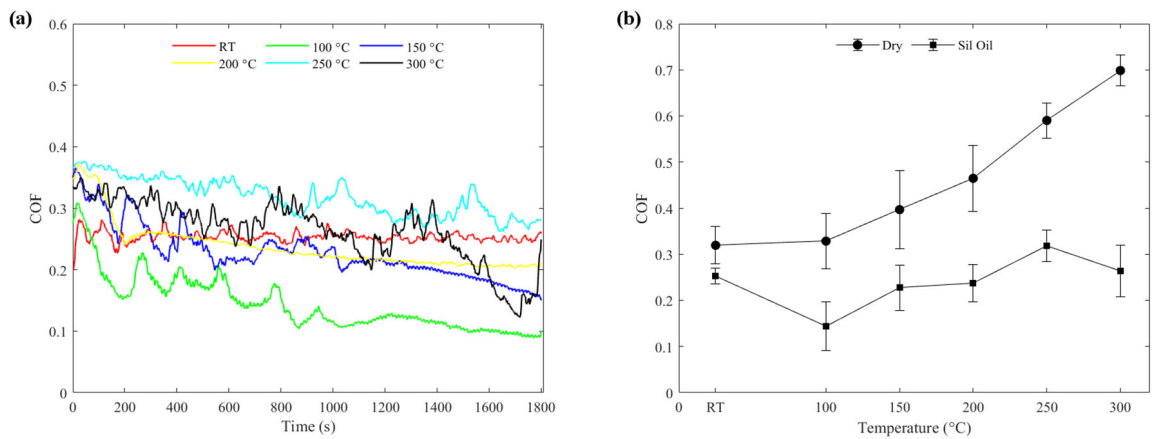

**Fig. 8** (a) Measured friction coefficient under the lubrication conditions, and (b) Average friction coefficient for both the dry and lubricated conditions.

Specific wear rate on the AZ31B disc samples was calculated by optically measuring the wear track, as shown in Fig. 9. Compared with the dry sliding tests, the wear tracks after lubricated tribo-tests become flattened with a considerable reduction in the worn depth, e.g., 14.6 μm in the current study and 42.1 μm under dry sliding at 300 °C [17]. For lubricated tests, the maximum wear track depth slightly decreases from RT to 200 °C but inversely increases at high temperatures, i.e., 250 and 300 °C. As shown in Fig. 9 and Fig. 10a, obvious pileups can be noted along the edges of the wear track at high temperatures (i.e., 250 and 300 °C). Due to the enhanced ductility under these temperatures [32], this morphology may be related to the plastic deformation of disc material from the worn regime. Additionally, the observed shallow wavelets across the wear track indicate material transfer and accumulation, which lowers down the wear volume compared with the dry sliding [51]. It should be noted that such pileups and wavelets were not observed in dry sliding tests under high temperatures (300 and 400 °C). One possible reason is that the very low ductility of disc surface material after severe oxidation limits the material deformation, as evidenced by the wear debris with brittle edges and cracks [17]. Another is that successive sliding removes the accumulated material and leads to a larger wear volume.

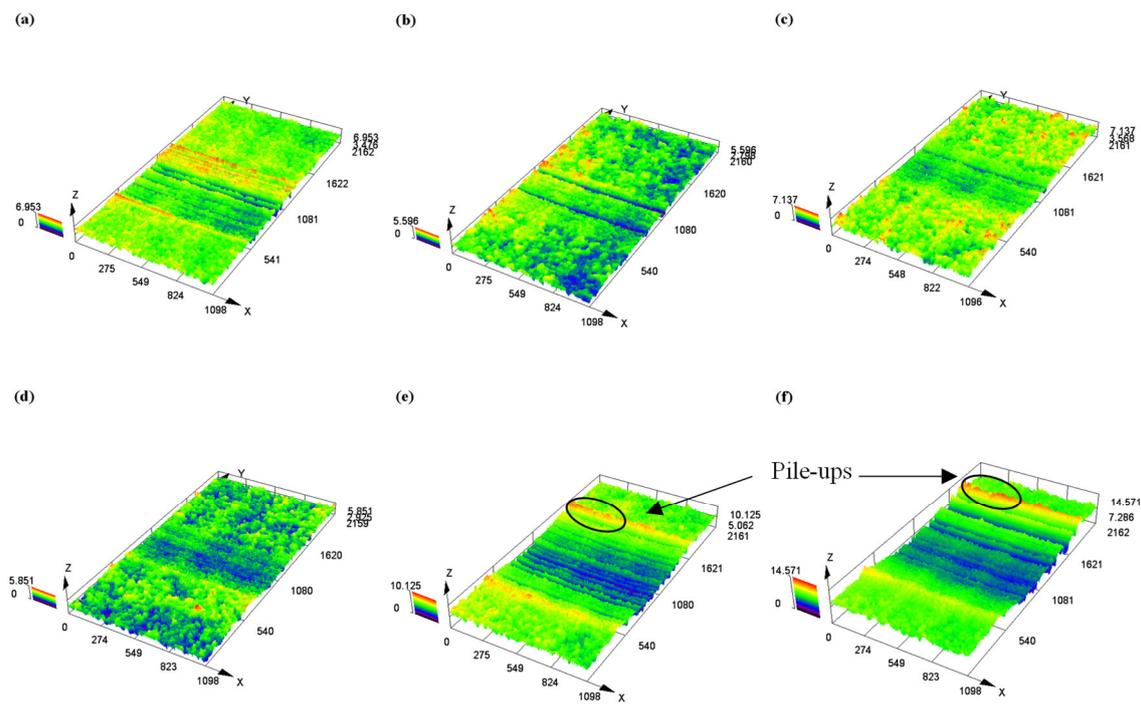

Fig. 9 3D profiles of wear tracks: (a) RT, (b) 100 °C, (c) 150 °C, (d) 200 °C, (e) 250 °C and (f) 300 °C.

Fig. 10b demonstrates the wear rate ($mm^3/N\,m$) of disc samples. For the dry sliding tests, the wear rate monotonically increases when the ambient temperature rises and a significant increase in the wear rate can be noted for a temperature above 250 °C. By contrast, the existence of silicone oil can effectively lower down the wear rate even at a high temperature, e.g., 300 °C. Below 200 °C, the silicone oil can enable a low wear rate. For the lubricated tests, there is a moderate increase in the wear rate at a temperature above 200 °C. This reveals that silicone oil can effectively improve the performance of tribology tests in terms of reducing the COF and wear rate over the temperature range considered in the current study.

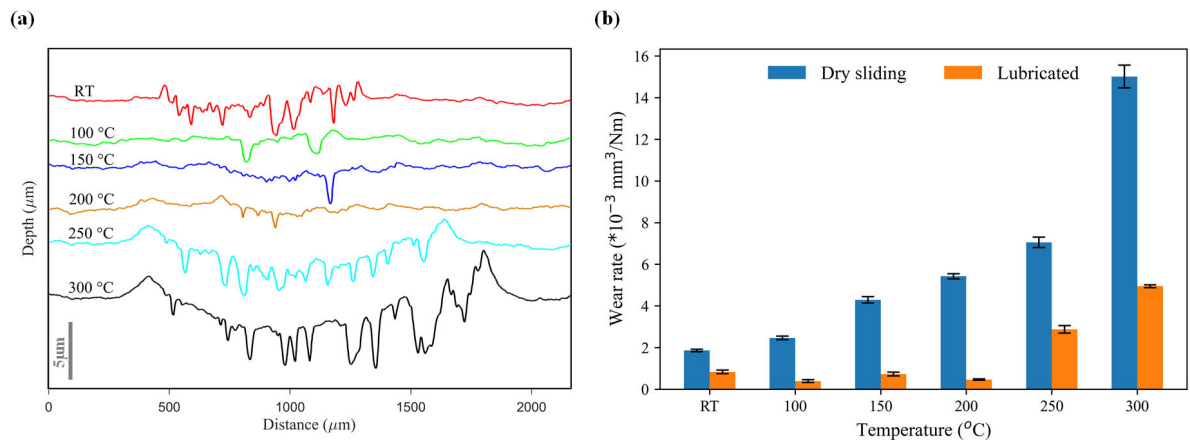

Fig. 10 (a) Cross-section of the worn profiles, and (b) wear rate of the disc samples.

## 3.3 Pin and disc surface after tests

Under the lubricated conditions, micro-scratches were observed on both the wear tracks and the pin surfaces, as shown in Figs. 11a and b. These scratches were possibly formed by the abrasive wear of asperity contact during sliding. Compared to the dry sliding contact, a darker colour can be noted on the scratches for the lubricated cases because of some silicone oil residue trapped inside the wear track during the successive sliding process [39]. At a high temperature of 300 °C, more cracks develop outward from the edge of the worn regime under lubrications. Additionally, some melted disc material was transferred to the pin surface, indicating adhesive wear, as shown in Fig. 11b. More pin and disc surfaces after tests under lubricated conditions are given in the appendix.

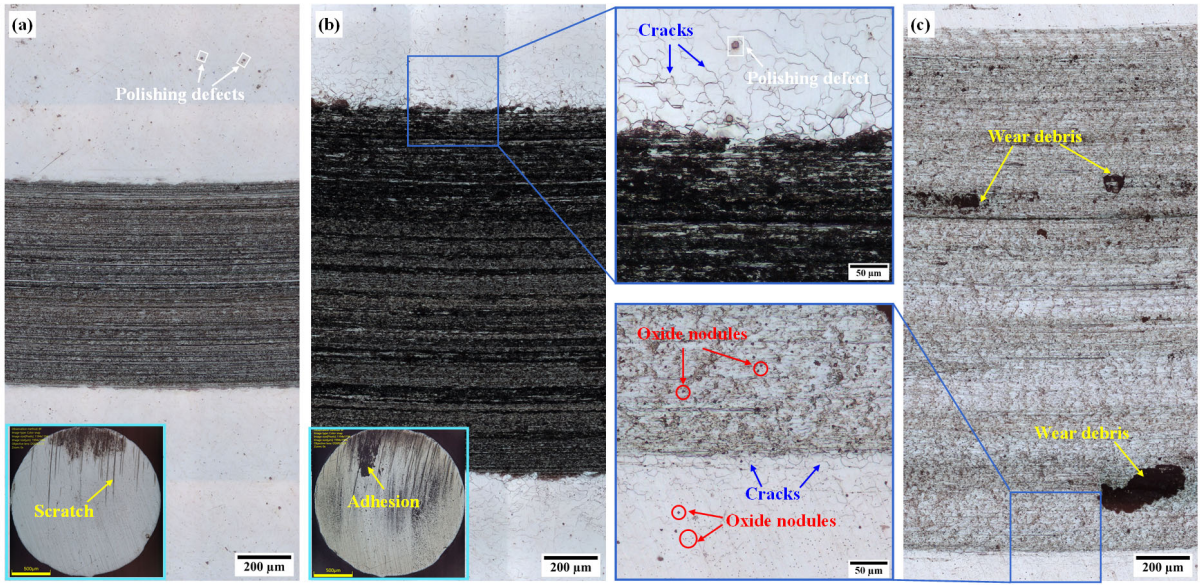

Fig. 11 Surface morphologies of pin and wear tracks, (a) lubricated test at RT, (b) lubricated test at 300 °C, and (c) dry sliding at 300 °C.

Considering the oxidation mechanism of Mg alloys and the stress-concentration phenomena in Fig. 6, such difference is mainly caused by the oxide inhibition of silicone oil, e.g., no oxide nodules observed both inside and outside the wear track at 300 °C. For the dry sliding tests at 250 °C, cauliflower-like oxide nodules can be noted inside the wear track but not in the non-sliding zone (see Figs. 12a and c). This is because the friction-induced heat brings about a higher temperature in the sliding zone where the oxide nodules more likely occur. For a sufficiently high temperature, e.g., 300 °C, oxide nodules can be seen both inside and outside the wear track, as shown in Figs. 12b and d. Typically, the oxide nodules lead to the formation of large wear debris which can be trapped on the wear track, as shown in Fig. 11c, leading to a larger fluctuation of COF during successive sliding.

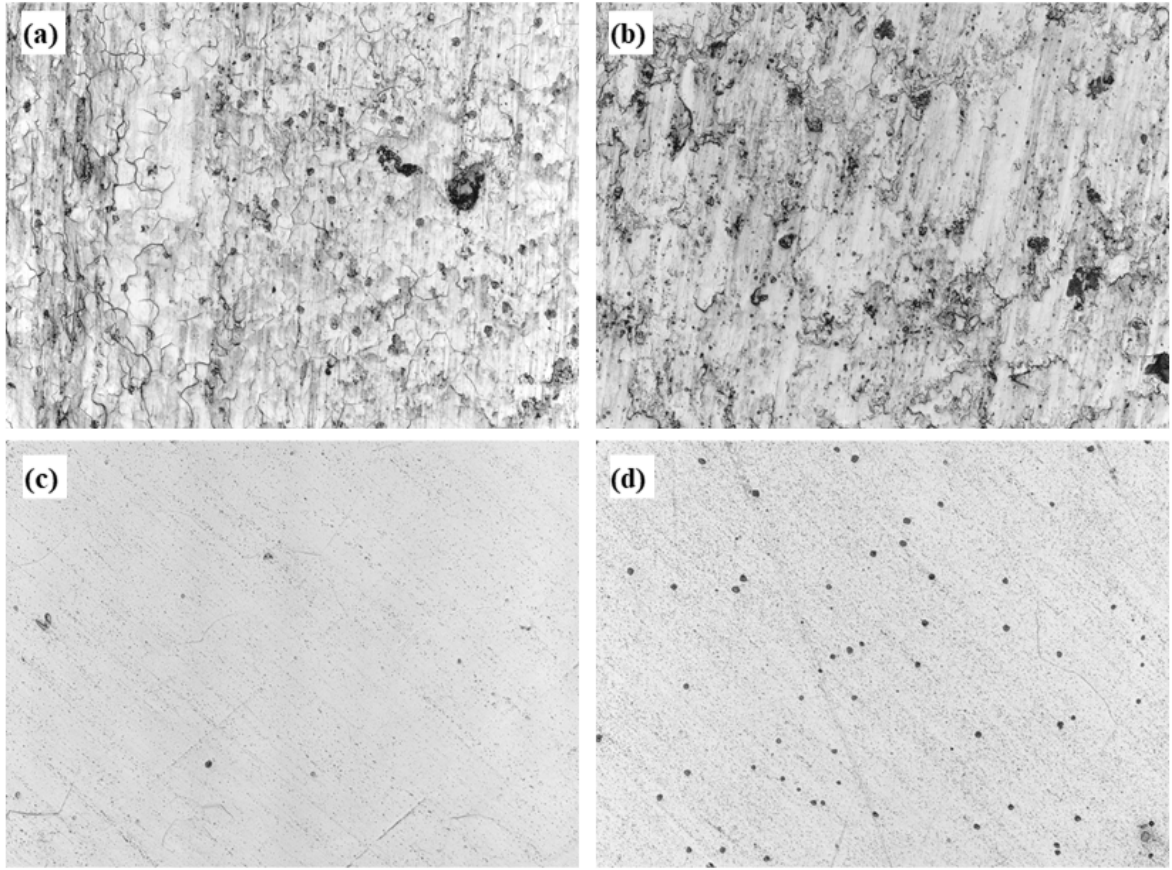

Fig. 12 Surface morphologies of wear tracks under dry sliding (a) sliding zone 250 ºC, (b) sliding zone 300 ºC, (c) non-sliding zone 250 ºC, and (d) non-sliding zone 300 ºC.

### 3.4 Wear mechanism

To further investigate the wear mechanism of AZ31B under the lubrication of silicone oil, the surface morphologies of the worn regime were characterized by a scanning electron microscope (SEM). At an ambient temperature below 100 ºC, some relatively flat micro ploughs can be noted in the wear tracks, as shown in Figs. 13a and b, indicating an abrasion-dominated wear behaviour. There are some unique flake-like areas for the temperature at 100 ºC. These relatively flat ploughs usually result in a stable sliding process, as reflected by the stable COF in Fig. 8a. The existence of these flat areas brings about a reduction in the roughness of the contact surface, resulting in the lowest friction coefficient at this temperature. At 150 ºC, cracks, approximately perpendicular to the sliding direction, were discovered in the wear track, which are typically linked to fatigue wear, as shown in Fig. 14a. A possible explanation for this is that repeated sliding processes led to a large strain accumulation [52]. Cracks were also found on the wear track at 200 °C. Additionally, deeper micro ploughs formed at the worn regime brought about an obvious fluctuation in COF, as shown in Fig. 14b. With a further increase in the ambient temperature, e.g., above 250 °C, the significantly deepened and widened ploughs (see Fig. 13c), and the enhanced thermal softening effects on the material both lead to a substantial rise in the COF (see Fig. 8a) and the material wear rate (see Fig. 10b). Moreover, the plastic deformation of material becomes obvious due to the thermal softening effects. Hence, the material in the ploughing regime experiences significant plastic deformation. The combined effects of pull-out material during the sliding and the lubrication at a high temperature, e.g., 300 °C, eventually result in some blunt protrusions along the sliding direction The observed galling on the surface indicates that the adhesion effects also become important.

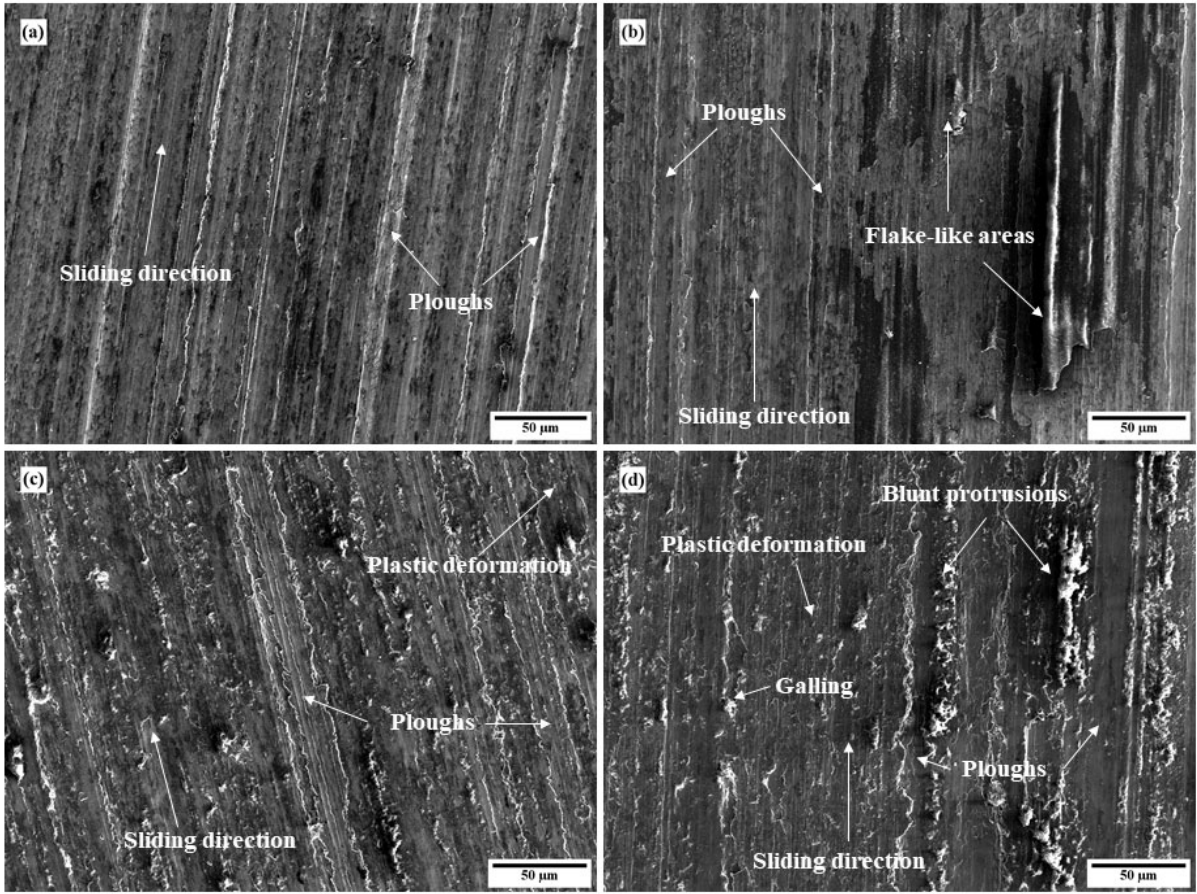

Fig. 13 SEM images of wear tracks under lubrication (a) RT, (b) 100 ºC, (c) 250 ºC and (d) 300 ºC.

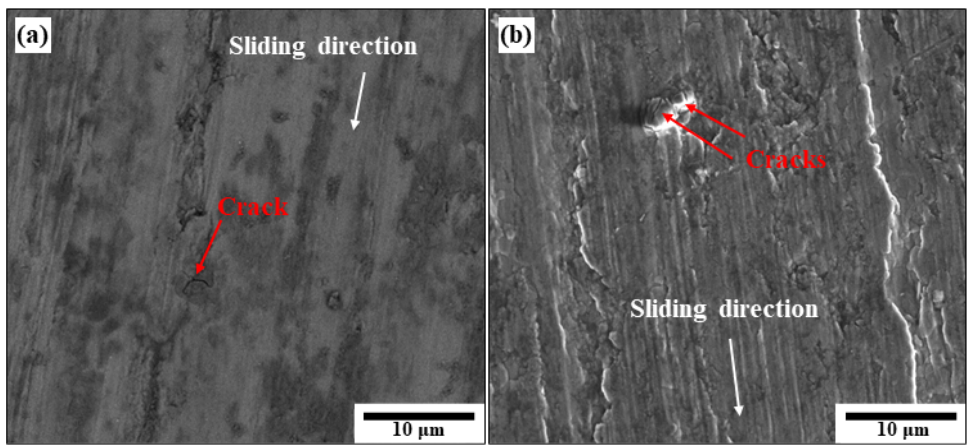

Fig. 14 High magnification SEM images of the wear tracks at (a) 150 ºC and (b) 200 ºC, respectively.

Under the dry sliding contact, more obvious ploughs along the sliding direction have been found in the current study, which is consistent with the available investigations [12, 17]. For a low temperature, clear ploughs with tiny debris can be easily identified on the surface, as shown in Figs. 15a and b. At an elevated temperature above 250 ºC, there is obvious plastic deformation. A deep galling induced by relatively large debris can be noted as well because the surface protective oxide layer has been destroyed after the initial generation of ploughs. Due to the friction-induced temperature rise in the contact area of the counterparts, oxide nodules are formed on the surface, as shown in Fig. 16, which further bring about the formation of a nonprotective oxide film. This undermines the wear resistance of the alloy and leads to larger adhesive debris pulled out from the disc surface, as shown in Fig. 15c. With the combined effects of material softening and sliding, the generated wear debris and pull-out material result in severe galling wear in the successive sliding process [53, 54]. A further increase in ambient temperature, i.e., 300 ºC, tends to enhance the formation of oxide nodules, which aggravates the contact conditions, as shown in Fig. 15d. These findings confirm those of earlier studies, which showed that the dominant wear mechanism at RT under dry sliding is abrasive wear. As the temperature increased, the main wear mechanism will transit to fatigue and adhesive wear and then to severe wear with the combination of abrasion, adhesion, oxidation, surface melting and fatigue [17].

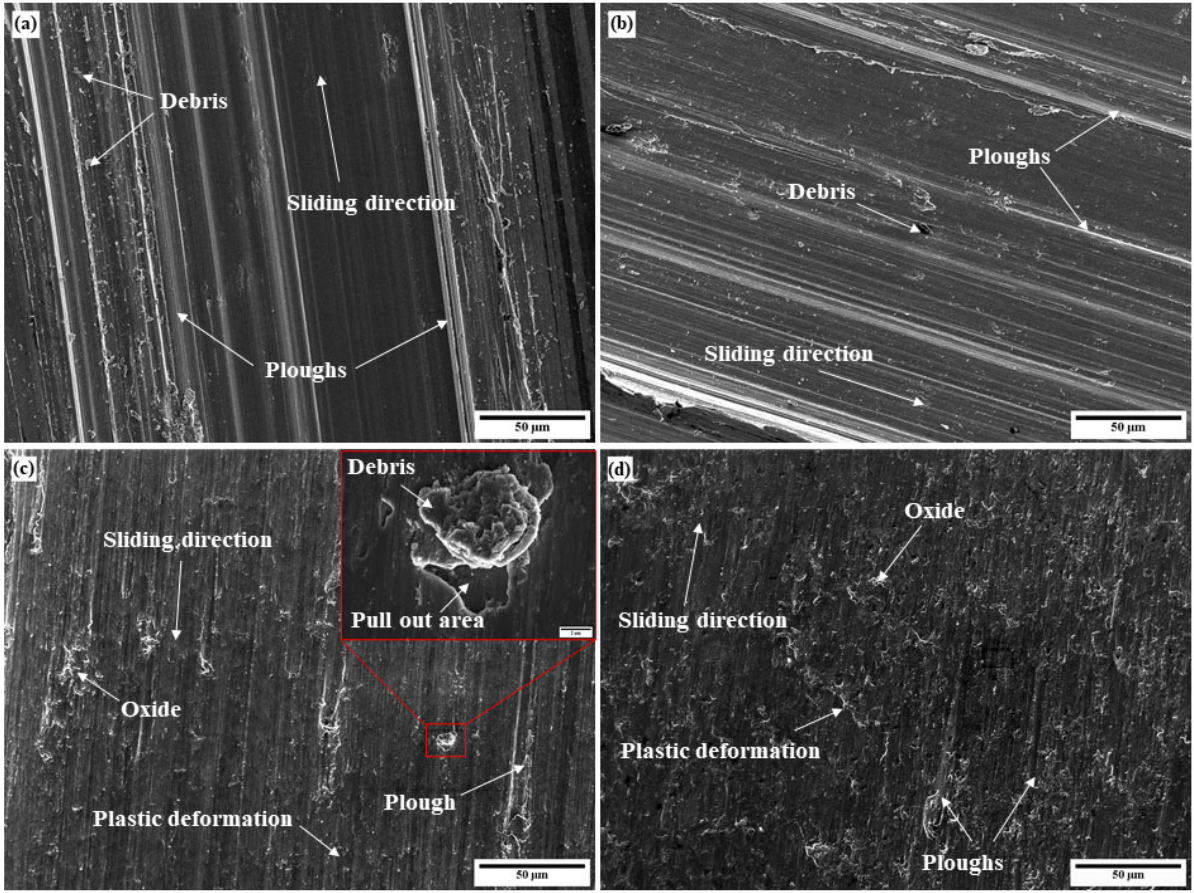

Fig. 15 SEM images of wear tracks under dry sliding (a) RT, (b) 100 ºC, (c) 250 ºC and (d) 300 ºC.

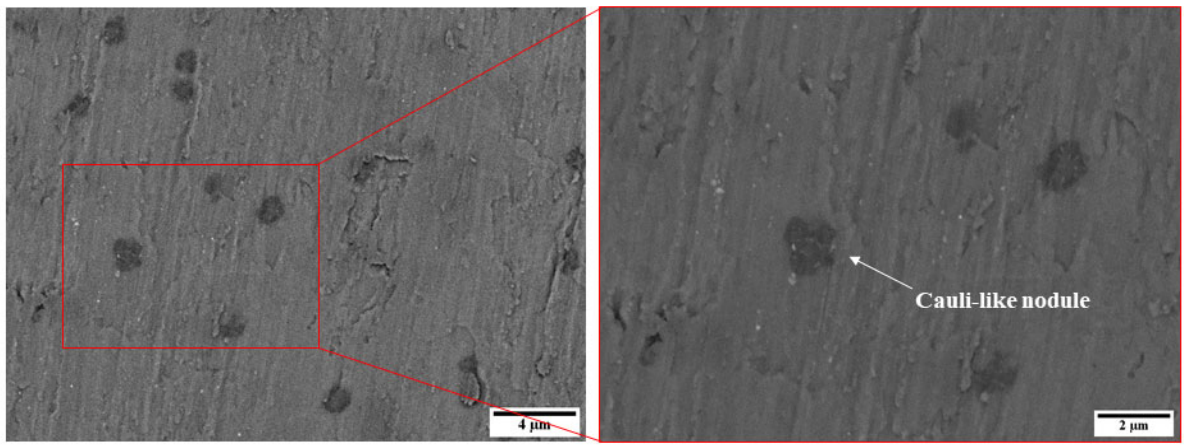

Fig. 16 SEM images of the wear track with oxide nodules after dry sliding test at 250 ºC.

Compared to the dry sliding at RT, the applied silicone oil shares a fraction of interface load and provides a boundary layer with a lower shear resistance, leading to a lower friction coefficient and wear rate. At a higher temperature of 100 °C under lubrication, the unique flake-like areas formed by the combined effects of operation conditions result in the lowest friction coefficient and wear rate. The surface morphologies between the dry sliding and the lubrication become different when the ambient temperature is above 250 °C. Morphologies under lubricated conditions show plasticity and adhesion with some blunt protrusions. Typically, the silicone oil inhibits the formation of nonprotective oxide nodules on the disc surfaces, lowers down the wear rate, and reduces galling. In contrast, the disc surface morphology under dry sliding experiences a strong effect of oxidation, adhesion, and fatigue [17].

**Conclusions**

This work has experimentally studied the oxidation inhibition and lubrication performance of silicone oil in the contact sliding of AZ31B Mg alloy at elevated temperatures. Based on the analysis of experimental results, the following conclusions can be drawn:

1. The silicone oil can prevent the exposure of Mg alloy to oxygen in the air, eliminate the formation of oxide films with oxide nodules in the alloy surfaces at elevated temperatures. However, cracks can still propagate in the AZ31B Mg alloy surface.

2. The silicone oil makes the coefficient of friction lower and more stable. At 100 °C, The coefficient reached its minimum. The oxidation inhibition effect of silicone oil prevents oxide nodules and reduces surface defects with shallower ploughs, which in turn leads to the reduction of wear rate and wear depth.

3. The morphologies of worn surfaces and tribological characteristics subjected to lubricated and dry sliding are significantly different. The lubricated wear track shows flake-like areas at 100 ºC, leading to reduced surface roughness and low friction coefficient. The morphologies under lubricated conditions show plastic deformation and adhesion with some blunt protrusions and pileups, indicating a combined wear mechanism of abrasion, fatigue and adhesion. However, the disc surface morphologies under dry sliding show a combined effect of severe oxidation, adhesion, surface melting and fatigue.

**Acknowledgement**

This research forms part of the Baosteel Australia Research and Development Centre (BAJC) portfolio of projects and has received financial support from the Centre through Project BA18002. This work has also been financially supported by the Chinese Guangdong Specific Discipline Project 2020ZDZX2006 and Shenzhen Key Laboratory of Cross-Scale Manufacturing Mechanics Project ZDSYS20200810171201007. The first author has been financially supported by the TFS of UNSW Sydney.

**Appendix**

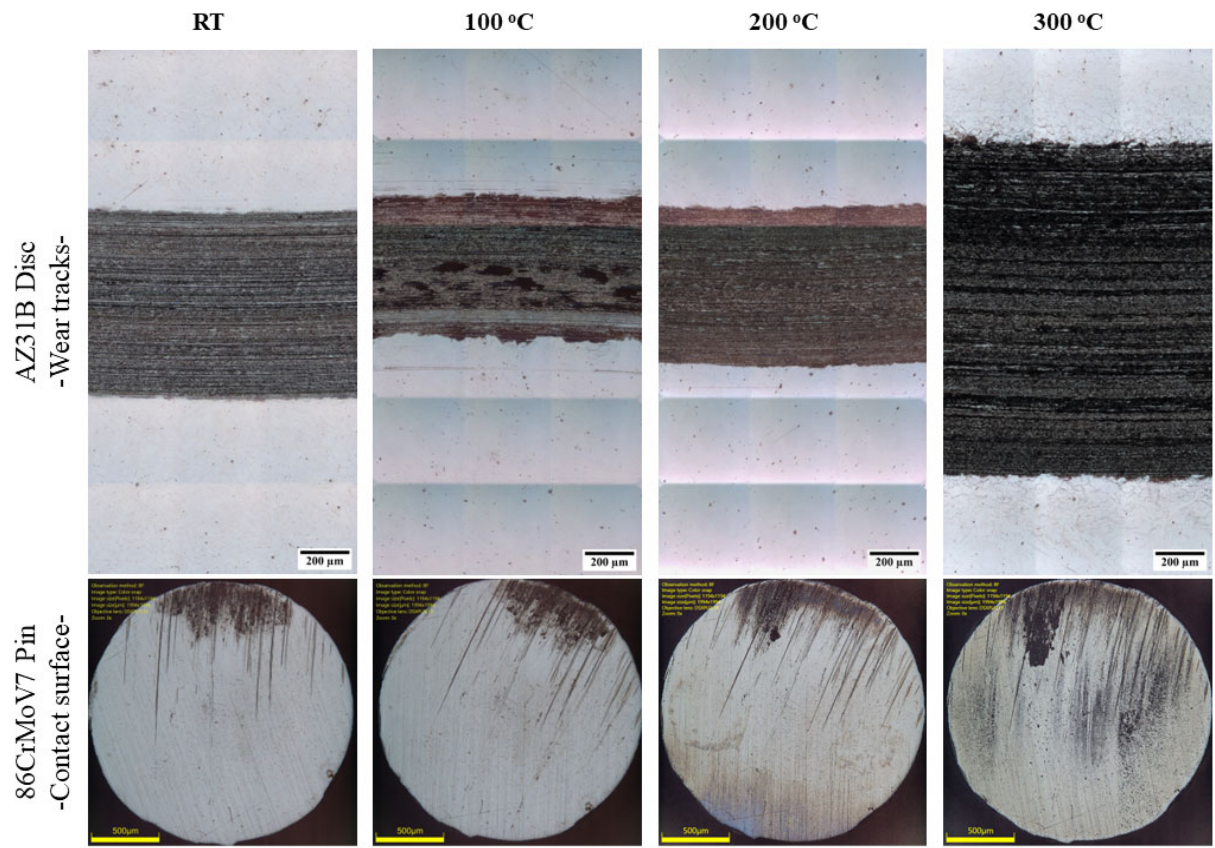

Fig. A1 Observations of wear tracks and pin surfaces after tests under lubricated conditions.